\documentstyle[epsfig]{aipproc}

\renewcommand{\baselinestretch}{1.01}
\newcommand{\be}{\begin{equation}}
\newcommand{\ee}{\end{equation}}

\newcommand{\bea}{\begin{eqnarray}}
\newcommand{\eea}{\end{eqnarray}}
\newcommand{\bean}{\begin{eqnarray*}}
\newcommand{\eean}{\end{eqnarray*}}

\def\se{\vskip3pt plus1pt minus1pt\setbox0=\hbox to\hsize\bgroup\hss
        \vrule width.5pt
        \vbox\bgroup \hrule width \hsize height.5pt
        \vskip3pt\hbox to\hsize\bgroup\hss\vbox\bgroup\advance\hsize by-9pt
        \columnwidth\hsize\small}
\def\ee{\par\egroup\hss\egroup\vskip3pt\hrule width\hsize height.5pt\egroup
        \vrule width.5pt\hss\egroup
        \box0 \vskip3pt plus1pt minus1pt}
\begin{document}
\title{CMB Signatures of a Primordial Magnetic Field}
\author{ 
Tina Kahniashvili$^{1,2\dagger}$, 
Arthur Kosowsky$^{2,\ddagger}$, 
Andrew Mack$^{2,\ddagger}$,
and Ruth Durrer$^{3, \dagger}$}
\address{
$^1$Center for Plasma Astrophysics, Abastumani 
Astrophysical Observatory, Kazbegi Ave.~2a, 380060 Tbilisi, Georgia \\
$^2$Dept.\ of Physics and Astronomy, 
Rutgers University, 136 Frelinghuysen Road, Piscataway, NJ 08854 USA\\
$^3$Universit\'e de Gen\`eve, D\'epartement de Physique Th\'eorique,
Switzerland\\
$^{\dagger}$tinatin,\,durrer@amorgos.unige.ch, $^{\ddagger}$kosowsky,\,andymack@physics.rutgers.edu}

\maketitle

\begin{abstract}
A primordial stochastic magnetic field will induce temperature and
polarization fluctuations in the cosmic microwave background. 
We outline a calculation of the resulting fluctuation power spectra
and present numerical results.
\end{abstract}

\section{Introduction}

The presence of a magnetic field in the early universe influences the
evolution of metric perturbations, and as a result, might leave
observable traces in the cosmic microwave background (CMB).  During
the past few years the gravitational effects of a homogeneous magnetic
field were presented by several authors; for a detailed review see
\cite{grasso} and references therein.  More realistic is the case of a
stochastic magnetic field, because any causal generation mechanism
results in stochastic fields.  Some estimates of CMB temperature power
spectra from density perturbations induced by a stochastic magnetic
field are given in \cite{lemoine} and numerical simulations are
presented in \cite{koh}. The vorticity induced by a stochastic
magnetic field was studied in \cite{SB} and CMB temperature
fluctuations were obtained in the case of a single $k$
mode. The gravitational waves generated by tangled magnetic fields and
the resulting tensor CMB temperature fluctuations spectra are given in
\cite{DFK}. 

Here we outline a comprehensive analytic calculation of temperature and
polarization power spectra due to all types of metric perturbations
arising from a primordial stochastic magnetic field, which we assume
to be statistically homogeneous and isotropic and described by a
power law. Detailed results will be presented elsewhere.

\section{Metric Perturbations Induced by Magnetic Fields}

The energy density of the magnetic field is treated as a 
first order perturbation, which thus does not alter the
background cosmology.
Prior to decoupling, the conductivity of the primordial
plasma can be assumed to be infinite, which in the comoving
frame implies the ``frozen-in'' condition 
${\mathbf E} = - {\mathbf v} \times {\mathbf B}$,
where ${\mathbf v}$ is the plasma velocity and ${\mathbf E}$ is
the electric field induced by plasma motions. Infinite
conductivity leads to a vanishing electric field in linear
perturbations theory ($v \ll 1$) and permits a split of the
spatial structure and time dependence of the magnetic field. As the
universe expands, magnetic field lines are simply conformally
diluted due to flux conservation: 
${\mathbf B}(\eta,{\mathbf x})= {\mathbf B}(\eta_0,{\mathbf x})/a^2$,   
where $a$ is the scale factor.

We assume that magnetic field is distributed homogeneously and
isotropically  with a power law power spectrum
\cite{DFK}:\
\begin{equation}
\langle B_i({\mathbf k})B^{*}_j({\mathbf k'})\rangle
=(2\pi)^5 \frac{\lambda^{n+3}}{\Gamma(\frac{n+3}{2})} B^{2}_\lambda
(\delta_{ij}-\hat{k}_i\hat{k}_j)
 k^n \delta({\mathbf k-k'}), ~~ k<k_D.
\end{equation}
The spectrum vanishes for all scales smaller than a damping scale
$k>k_D$; we require the spectral index $n>-3$ so as not to overproduce
superhorizon coherent fields. $B_\lambda$ is the rms magnetic field
strength today smoothed over a co-moving length scale $\lambda$.

The electromagnetic stress-energy tensor may be geometrically
decomposed into scalar (density), vector (vorticity) and tensor
(gravitational wave) perturbation modes:
$\tau^{(B)}_{ij}=\Pi_{ij}^{(S)}+\Pi_{ij}^{(V)}+\Pi_{ij}^{(T)}$.  From
the tensor $\Pi_{ij}^{(S)}$ we can construct scalar $\Pi^{(S)}$ which
describes the density perturbations, and from the tensor $\Pi_{ij}^{(V)}$
a vector $\Pi^{(V)}_i$ which describes the vorticity perturbations.
Then $\Pi^{(S)}$ is proportional to the magnetic field energy density
$|\rho_B({\mathbf k}, \eta_0)| \simeq {3 \over 2} |\Pi^{(S)}({\mathbf
k}, \eta_0)|$, and $\Pi_{i}^{(V)}$ is proportional to the vortical
piece of the Lorentz
force, $L^{(V)}_i=k\Pi^{(V)}_i$.  The isotropic spectra $\Pi^{(S),
(V), (T)}$ appear as source terms for scalar, vector and
tensor perturbations (for details see \cite{DFK}, \cite{KKM1}).

To obtain the evolution equations for perturbations, we must 
derive the scalar, vector and tensor magnetic field correlation functions for
$\Pi^{(S)}$, $\Pi^{(V)}_i$ and $\Pi^{(T)}_{ij}$ as described
in \cite{DFK} and obtain the corresponding spectra as a functions
of $n$, $B_\lambda$ and $k_D$. Terms induced
by the magnetic field (e.g.\ Lorentz force, magnetic field anisotropic 
stresses) appear only in the dynamical equations describing baryons. 
We use notational conventions of Ref.~\cite{cmbform1} and employ the
Newtonian gauge; $a$ is the scale factor, $\eta$ is conformal time,
and 0 subscripts refer to the present time.

{\it Scalar perturbations:} Since we assume that the background
spacetime is unperturbed and linear order perturbations arise only
from the magnetic field, the stress energy of the magnetic
field is not compensated by
anisotropic stress of the plasma and we have non-zero
initial conditions for gravitational potentials.
Such initial conditions might arise from an inflation-like
process; the compensated case, arising from causal processes like
phase transitions, has more complicated dynamical effects, but our
estimates should still be of the correct order of magnitude.
We limit the spectral index of the magnetic field to
$n < 2$ \cite{DFK}; the case $n > 2$ corresponds to causally
generated fields, which also must be compensated.

The scalar constraint equations from the Einstein equations
are two Poisson equations,
\begin{equation}
k^2 \Phi = 4 \pi G a^2 [\rho_f \delta_f +
\rho_B],\qquad\qquad 
k^2 (\Psi + \Phi) = - 12 \pi G a^2 \Pi^{(S)},
\label{scalarconstraints}
\end{equation}
where $\rho_f \delta_f$ is the total fluid perturbation. 
The dynamical Einstein equations determine the evolution of the
photon and baryon densities (see also \cite{Adams}): 
\begin{equation}
\dot \delta_\gamma = - {4 \over 3} (k v^{(S)}_\gamma \!+ \! 3 {\dot
\Phi}),\quad   \dot\delta_{\rm cdm} = -(k v^{(S)}_{\rm cdm}\! + \! 3 {\dot
\Phi}),\quad
\dot \delta_b = - (k v^{(S)}_b \!+ \! 3 {\dot \Phi}) \! -\! 
3 {\dot a \over a} {\Pi^{(S)} \over \rho_b}.
\label{scalardynamics}
\end{equation}
Using the zero-order approximation for tight coupling regime, 
$v_b \simeq v_\gamma
\simeq v^{(S)}$, we obtain the following
equation of motion \cite{KKM1} from momentum conservation
equations for baryons and photons:
\begin{equation}
\dot v^{(S)} - {k \over 4} \delta_\gamma  -
k \Psi +
{k \Pi^{(S)} \over \rho_\gamma+p_\gamma}=0.
\label{vdot}
\end{equation}
Using the equations of motion, it is possible to show that
the two terms on the right side of the first Poisson equation
are roughly equal. An approximate solution for the gravitational
potential is then
\begin{equation}
\Phi(\eta) \simeq {12 \pi G \Pi^{(S)}(\eta_0, k) \over k^2
a^2} . 
\label{phisolution}
\end{equation}

{\it Vector perturbations:} The vector Einstein equation which
describes the evolution of the vector
potential sourced by a stochastic magnetic field is
\begin{eqnarray}
\dot{V}_i+2{\dot a\over a}V_i = -16\pi Ga^2\Pi^{(V)}_i(\eta,{\mathbf k})/k.
\label{eq:V-Einstein-1}
\end{eqnarray}

Another constraint equation relates the vector
potential to the vorticity ${\bf\Omega}$:
\begin{equation}
-k^2 V_i = 16\pi Ga^2(\rho+p)\Omega_i. 
\label{eq:V-Einstein-2}
\end{equation}
Introducing the Lorentz force term
into the baryon Euler equation, neglecting baryon inertia, and
solving the baryon momentum conservation equation in the
tight-coupling approximation, we obtain an approximate solution for
the vorticity,
\begin{equation}
{\mathbf\Omega}(\eta,{\mathbf k})
\simeq \frac{k{\mathbf\Pi}^{(V)}(\eta_0,{\mathbf k})\eta}{\rho_
{\gamma0}+p_{\gamma0}}. 
\label{eq:V-vorticity-soln}
\end{equation}

{\it Tensor perturbations:} The tensor Einstein equation that
describes the evolution of gravitational waves
sourced by a stochastic magnetic field is \cite{lemoine,DFK}
\begin{eqnarray}
\ddot{H}_{ij}+2\frac{\dot{a}}{a}\dot{H}_{ij}+k^2H_{ij}=
8\pi Ga^2\Pi^{(T)}_{ij}(\eta,{\mathbf k}).
\end{eqnarray}
The tensor case has no additional constraint equations.
As in the case of vector perturbations,
we neglect the tensor anisotropic stress of the plasma,
which is in general negligible.
A Green function technique
gives an approximate solution for $\eta > \eta_{\rm eq}$ \cite{DFK}
\begin{eqnarray}
\dot{H}(\eta,k)\simeq 4\pi G\eta^2_0z_{\rm eq}\ln\left(\frac{z_{\rm
in}}{z_{\rm eq}}\right)k\Pi^{(T)}
(\eta_0,k)\frac{j_2(k\eta)}{k\eta},
\end{eqnarray}
where $H(\eta, k)$ and $\Pi^{(T)}(\eta_0,k)$
are isotropic correlation functions of the tensors $H_{ij}$ and $\Pi^{(T)}_
{ij}$ respectively, $z_{\rm in}$ is the initial redshift at which
the magnetic field is generated, and $z_{\rm eq}$ is the redshift at
matter-radiation equality.

\section{CMB Anisotropies}

The solutions for the metric perturbations presented above can be used
to compute the power spectra of temperature and polarization
anisotropies in the CMB. The contributions from scalar, vector, and
tensor perturbations are all uncorrelated. For all three cases,
the temperature fluctuations arise simply from the Sachs-Wolfe
effect, while the polarization fluctuations are sourced by these
temperature fluctuations. For the scalar case, the temperature
perturbation is just the familiar $-\Phi/3$ at decoupling. The vector
and tensor cases are slightly less familiar; expressions are given
in Ref.~\cite{Durrer} for vector perturbations 
and Ref.~\cite{DFK} for tensor perturbations. 

We have derived
analytic approximations to the  angular power spectra of temperature
and polarization for all
three types of metric perturbation, 
neglecting small-scale damping effects due to photon diffusion 
or magnetic damping. As examples, for {\em scalar
perturbations} with $n< -3/2$, the temperature power spectrum is
\begin{equation}
\ell^2C_\ell^{TT} \simeq {(2\pi)^5\over 3}{G^2 \eta_0^4 \over a_{\rm dec}^4}
{\scriptsize 2^{2n+1} n\over (n+3)(2n+3)}
{\scriptsize \Gamma(-2n)\over\Gamma^2\left(-n+{1\over 2}\right)
\Gamma^2\left({n+3\over 2}\right)}
\left(\lambda\over\eta_0\right)^{2n+6} \ell^{2n+2} B_\lambda^4,
\label{scalarClT}
\end{equation}
while the power spectrum for the polarization-temperature cross
correlation is
\begin{equation}
\ell^2C_\ell^{TE} \simeq -{8\pi^4\over 9}{ \scriptsize 
G \eta_0^2 \over \rho_{\gamma 0}\! + \! p_{\gamma 0}}
{\scriptsize 2^{2n\! -\! 1} n\over (n\! +\! 3)(2n\! +\! 3)}
{\scriptsize  \Gamma(-2n\!-\! 2)\over\Gamma^2\left(-n\!-\!{1\over 2}\right)
\Gamma^2\left({n\! +\! 3\over 2}\right)}
\left(\lambda \ell\over\eta_0\right)^{2n+6} 
\left({\eta_0\over \eta_{\rm dec}}\right)^2 B_\lambda^4.
\label{scalarClTE}
\end{equation}
Complete expressions for all cases will be presented elsewhere;
only the temperature power spectrum for tensors \cite{DFK}
and a partial calculation of the temperature power spectrum for
vectors \cite{SB} have been worked out previously.

The total power spectra for $n=-2$ are shown in the Figure.
For the smoothing scale we take $\lambda=0.1$ Mpc, which corresponds to
the maximum comoving
length scale of galaxies (as in \cite{DFK}); 
for the magnetic field damping scale we take the 
Alfven cut-off scale \cite{SB,DFK},  
which is roughly $k_D \simeq 4.5 h$ Mpc$^{-1}$. Photon diffusion 
and magnetic
damping will give an exponential cutoff to 
this fluctuation power spectra above a characteristic value of $\ell$.
Note that for $n> -2$, the polarization power spectra are actually 
larger than the temperature power spectrum for $l>50$.

\begin{figure}[t!]
\centerline{\epsfig{file=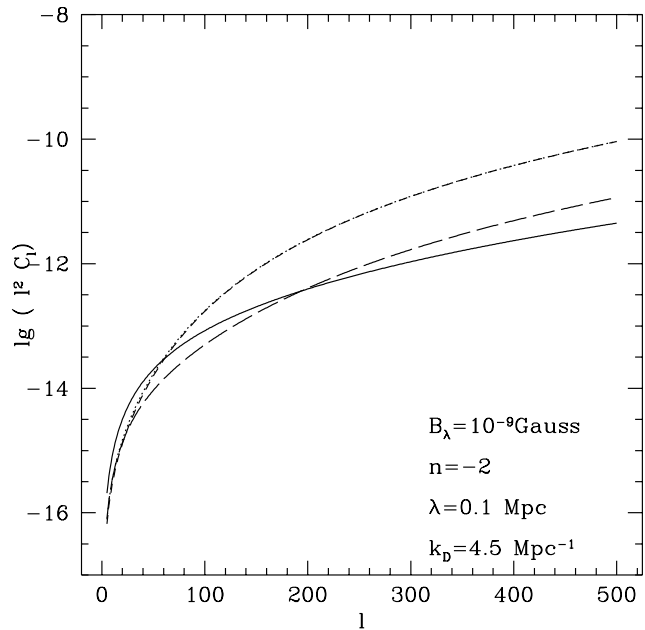, height=3.5in,width=3.5in}}
\caption{The total power spectra $\ell^2 C_\ell$ vs. $\ell$ 
for $n=-2$ are shown:  $\ell^2 C_\ell^{TT}$ - solid line, 
$\ell^2 C_\ell^{EE}$ - dotted line, $\ell^2 C_\ell^{BB}$ - short dashed line, 
$\ell^2 C_\ell^{TE}$ long-dashed line.}
\vspace*{10pt}
\end{figure}

The magnetic field spectral index is confined to the range  
$-3 < n < 2$.
As shown in \cite{DFK,KKM1}, no spectral index will produce 
a scale-invariant CMB power spectrum ($\ell^2 C_\ell$ constant) for scalar 
temperature and polarization anisotropies 
while $n \rightarrow -3 $ results in 
a scale-invariant cross correlation spectrum $\ell^2 C_\ell^{TE}$. 
The vector perturbations cannot produce scale-invariant spectra for 
allowed values of $n$; for tensor 
perturbations both, the temperature  and polarization spectra  $\ell^2
C_\ell$ are scale-invariant for $n \rightarrow -3$.   
Scalar perturbations are always subdominant if compared to vector and
the dominant tensor perturbations.

The CMB power spectra due to a magnetic field vary in amplitude like
the square of the energy density perturbations, or as $B^4$. It will
thus be impractical to obtain upper limits on magnetic field strengths
which are significantly more stringent than around $10^{-9}$ Gauss;
limits improve as $n$ decreases. On
the other hand, if primordial fields are present at this level, the 
combined signature in the various microwave background power spectra
will give an unmistakable signal. We also note that the vector
and tensor perturbations generated by magnetic fields are one of the
few cosmological sources of ``B'' polarization \cite{kkt}, along with
primordial tensor perturbations and gravitational lensing of the
microwave background. Stochastic magnetic fields also will induce
Faraday rotation in the microwave background polarization, providing
an additional ``B'' polarization signal; we are currently considering
this question.

\smallskip

{\it Acknowledgments:}
This work has been supported by the COBASE program of the
U.S.\ National Research Council and by the
NASA Theory Program through grant NAG5-7015. 
T.K. acknowledges a grant from the Swiss NSF for participation in CAPP2000.
A.K. is a Cotrell Scholar of the Research Corporation.

\end{document}